\documentclass[journal, final, onecolumn,12pt]{IEEEtran}
\makeatletter
\def\ps@headings{%
\def\@oddhead{\mbox{}\scriptsize\rightmark \hfil \thepage}%
\def\@evenhead{\scriptsize\thepage \hfil \leftmark\mbox{}}%
\def\@oddfoot{}%
\def\@evenfoot{}}
\makeatother
\usepackage{cite}
\usepackage{subfigure}
\usepackage{graphicx}
\usepackage{mathrsfs}
\usepackage{amsmath}
\usepackage{amsfonts}
\usepackage{amssymb}
\usepackage{paralist}

\newtheorem{theorem}{{Theorem}}

\newtheorem{corollary}[theorem]{{Corollary}}
\newtheorem{definition}{{Definition}}

\newcommand{\mb}{\mathbf}
\newcommand{\qed}{\hspace*{\fill} $\Box$ \\}

\allowdisplaybreaks

\begin{document}

\title{Directed Percolation in Wireless Networks with Interference and Noise}

\author{Zhenning~Kong,~\IEEEmembership{Student Member,~IEEE,}
        Edmund~M. Yeh,~\IEEEmembership{Member,~IEEE,}
\thanks{This research is supported in part by National Science Foundation (NSF) grant CNS-0716335, and
Army Research Office (ARO) grant W911NF-07-1-0524.}
\thanks{Z. Kong and Edmund~M. Yeh are with the Department of Electrical Engineering, Yale University
(email: zhenning.kong@yale.edu, edmund.yeh@yale.edu)}}

\markboth{Submitted to IEEE Transactions on Information Theory}{Submitted to IEEE
Transactions on Information Theory}

\maketitle

\begin{abstract}
Previous studies of connectivity in wireless networks have focused on undirected
geometric graphs. More sophisticated models such as
Signal-to-Interference-and-Noise-Ratio (SINR) model, however, usually leads to directed
graphs. In this paper, we study percolation processes in wireless networks modelled by
directed SINR graphs. We first investigate interference-free networks, where we define
four types of phase transitions and show that they take place at the same time. By
coupling the directed SINR graph with two other undirected SINR graphs, we further obtain
analytical upper and lower bounds on the critical density. Then, we show that with
interference, percolation in directed SINR graphs depends not only on the density but
also on the inverse system processing gain. We also provide bounds on the critical value
of the inverse system processing gain.
\end{abstract}

\baselineskip 20 pt

\section{Introduction}

The study of coverage, connectivity, and capacity in large-scale wireless networks from a
percolation perspective has attracted much attention recently~\cite{BoNrFrMe03,
FrBoCoBrMe05, DoFrTh05, DoBaTh05, DoFrMaMeTh06, FrDoTsTh07}. To intuitively understand
percolation processes in large-scale wireless networks, consider the following example.
Suppose a set of nodes are uniformly and independently distributed at random over an
area. All nodes have the same transmission radius, and two nodes within a transmission
radius of each other can communicate directly. At first, the nodes are distributed
according to a very small density. This results in isolation and no communication among
nodes. As the density increases, some clusters in which nodes can communicate with one
another directly or indirectly (via multi-hop relay) emerge, though the sizes of these
clusters are still small compared to the whole network. As the density continues to
increase, at some critical point a huge cluster containing a large portion of the nodes
forms. This phenomenon of a sudden and drastic change in the global structure is called a
\emph{phase transition}. The density at which the phase transition takes place is called
the \emph{critical density}. A fundamental result of continuum percolation concerns such
a phase transition effect whereby the macroscopic behavior of the system is very
different for densities below and above the critical density $\lambda_c$. For $\lambda <
\lambda_c$ (subcritical), the connected component containing the origin (or any other
node) contains a finite number of nodes almost surely. For $\lambda> \lambda_c$
(supercritical), the connected component containing the origin contains an infinite
number of nodes with a positive probability~\cite{Gi61, Gr99, MeRo96, Pe03, BoRi06}.

Most previous work employing continuum percolation for the study of large-scale wireless
networks have focused on geometric models in which a link exists between two nodes when
they are within each others' transmission radii. This simple model leads to an {\em
undirected} graph and considerably simplifies the resulting analysis. To reflect
realistic conditions in wireless networks, however, more sophisticated models for link
connectivity can be adopted. For instance, a widely-used model for wireless communication
channels is the Signal-to-Interference-and-Noise-Ratio (SINR) model \cite{Pr00, TsVi05}.
Here, the ability to decode the transmitted signal from node $i$ to node $j$ is
determined by the SINR\footnote{Note that the interference term in the conventional
definition of SINR is $\gamma\sum_{k\neq i}P_kL(d_{kj})$ rather than $\gamma\sum_{k\neq
i,j}P_kL(d_{kj})$\cite{Pr00, TsVi05}. The latter approach assumes that each node knows
its own transmitted signal and can subtract it from the received signal. In order to keep
consistent with previous work~\cite{DoBaTh05, DoFrMaMeTh06}, we follow this assumption in
this paper, though the results of this paper does not rely on this assumption.}
\begin{equation}\label{beta-ij}
\beta_{ij}=\frac{P_iL(d_{ij})}{N_0+\gamma \sum_{k\neq i,j} P_kL(d_{kj})}.
\end{equation}
where $P_i$ is the transmission power of node $i$, $d_{ij}$ is the distance between nodes
$i$ and $j$, and $N_0$ is the power of background noise. The parameter $\gamma$ is the
inverse of system processing gain.  It is equal to 1 in a narrowband system and smaller
than 1 in a broadband (e.g., CDMA) system. The signal attenuation $L(d_{ij})$ is a
function of distance $d_{ij}$.  Under the SINR model, the transmitted signal of node $i$
can be decoded at $j$ if and only if $\beta_{ij} \geq\beta$, where $\beta$ is some
threshold for decoding. In this case, a link $(i,j)$ is said to exist from $i$ to $j$.
Percolation in wireless networks under the SINR model has been studied for the {\em
undirected} case in~\cite{DoBaTh05, DoFrMaMeTh06}. Here, it is assumed that the
(undirected) link $(i,j)$ exists if and only if $\min \{\beta_{ij}, \beta_{ji}\} \geq
\beta$. Nevertheless, even if $\beta_{ij} \geq \beta$, $\beta_{ji} \geq \beta$ may not
hold and thus the link $(j,i)$ may not exist. Thus, the graph resulting from the SINR
model is in general {\em directed.}

Percolation processes in directed lattices have been well studied (see \cite{Gr99} and
references therein.) Recently, based on generating function methods, percolation has been
analyzed in directed scale-free random graphs~\cite{ScCoAvBaHa02}, and random graphs with
given degree distributions~\cite{NeStWa01}. Note, however, that lattices have regular
geometry, and both scale-free random graphs and random graphs with given degree
distributions lack the geometric constraints which exist in SINR graphs. Hence the
results and analytical methods in~\cite{Gr99, ScCoAvBaHa02, NeStWa01} are not directly
applicable for directed SINR graphs.

To understand how the directional nature of communication links affects the connectivity
of wireless networks, we first study percolation processes in the SINR model with $\gamma
= 0$ (interference-free directed graphs), where (directed) link $(i,j)$ exists if and
only if the distance between $i$ and $j$ is less than or equal to the transmission radius
associated with node $i$. In such directed graphs, a node has two types of links,
\emph{in-links}, which are the links pointing to the node from other nodes, and
\emph{out-links}, which are the links pointing out to other nodes. Indeed we can define
four types of components with respect to a given node $u$: \emph{in-component},
\emph{out-component}, \emph{weakly connected component} and \emph{strongly connected
component}. Corresponding to these four types of components, we can define four types of
phase transitions, and further define four corresponding critical densities. We will show
that all four critical densities are equal to a positive and finite value
$\overrightarrow{\lambda_c}(\mathcal{P})$, which depends on the power distribution at
each node. By coupling the directed SINR graph with two other types of undirected SINR
graphs and using cluster coefficient~\cite{KoYe07-1,KoYe07-2} and re-normalization
methods~\cite{MeRo96}, we further provide analytical upper and lower bounds on
$\overrightarrow{\lambda_c}(\mathcal{P})$.

Next, we show that with interference ($\gamma>0$), percolation in directed SINR graphs
depends not only on the density but also on the inverse system processing gain $\gamma$.
Indeed there exists a positive and finite critical value
$\overrightarrow{\gamma_c}(\lambda)$, such that the network is percolated only when
$\lambda>\overrightarrow{\lambda_c}(\mathcal{P})$ and
$\gamma>\overrightarrow{\gamma_c}(\lambda)$. Furthermore, for $\lambda$ sufficiently
large, $\overrightarrow{\gamma_c}(\lambda)=\Theta(\frac{1}{\lambda})$. The same results
have been obtained in~\cite{DoBaTh05, DoFrMaMeTh06} for undirected SINR graphs. Our
results indicate that the critical inverse system gain has the same asymptotical behavior
in directed and undirected SINR graphs.

The remainder of this paper is organized as follows. In Section II, we give the
definitions and assumptions for the directed SINR graph model. In Section III, we study
directed percolation in wireless networks without interference ($\gamma=0$). We formally
define four types of phase transitions and their corresponding critical densities, and
provide analytical lower and upper bounds for these densities. In Section IV, we
investigate directed percolation in wireless networks with interference ($\gamma>0$). In
Section V, we present simulation results on directed percolation in SINR graphs, and
finally, we conclude in Section VI.

\section{Network Model}

Although some of our results apply to $d$-dimensional graphs in general, we will focus on
the 2-dimensional case. Let $\|\cdot\|$ be the Euclidean norm, and $A=|{\cal A}|$ be the
area of ${\cal A}$.  Assume ${\mb X}_1, {\mb X}_2, ..., {\mb X}_n$ are i.i.d.
2-dimensional random variables with a common uniform density on a 2-dimensional box
${\cal A}=[0,\sqrt{n/\lambda}]^2$, where ${\mb X}_i$ denotes the random location of node
$i$ in $\mathbb{R}^2$. We assume that the transmission power $P_i$ are distributed i.i.d.
according to a probability distribution $f_P(p)$, $p\in [p_{min},p_{max}]$, where
$0<p_{min}\leq p_{max}<\infty$. This reflects heterogeneity of transmission powers in
real wireless networks. We further assume
\begin{itemize}
\item[(i)] $p_{min}\geq\beta N_0$; and \item [(ii)] $\mbox{Pr}\{P=p_{min}\}>0,
\mbox{Pr}\{P=p_{max}\}>0$.
\end{itemize}

In wireless networks under the SINR model, there is a directed link from node $i$ to node
$j$ if $\beta_{ij}\geq \beta$, and the link is bidirectional if and only if
$\min\{\beta_{ij}, \beta_{ji}\}\geq \beta$. Denote by
$\overrightarrow{G}(\mathcal{X}_n,\mathcal{P},\gamma)$ the ensemble of \emph{directed}
graphs induced by the SINR model. In order to show the percolation behavior in
$\overrightarrow{G}(\mathcal{X}_n,\mathcal{P},\gamma)$, we define two other types of
\emph{undirected} SINR graphs: the first is $G(\mathcal{X}_n,\mathcal{P},\gamma)$, where
there exists an undirected link between nodes $i$ and $j$ if and only if
$\min\{\beta_{ij}, \beta_{ji}\}\geq \beta$. The second is
$G'(\mathcal{X}_n,\mathcal{P},\gamma)$, where there exists an undirected link between
nodes $i$ and $j$ if and only if $\max\{\beta_{ij}, \beta_{ji}\}\geq\beta$. The model of
$G(\mathcal{X}_n,\mathcal{P},\gamma)$ was used in~\cite{DoBaTh05, DoFrMaMeTh06} as a
simplified physical model for wireless communication networks.

The sum $\sum_{k\neq j}L(d_{kj})$ is a random variable which depends on the locations of
all nodes in the network. The quantity
\begin{equation}\label{J-X-0}
J(\mathbf{x})\triangleq\sum_{i:\mathbf{X}_i\neq
\mathbf{x}}P_iL(||\mathbf{X}_i-\mathbf{x}||), \quad  \mathbf{x}\in \mathbb{R}^2.
\end{equation}
is called Poisson shot noise~\cite{Da71, DoBaTh05, DoFrMaMeTh06}. When $P_i$ is uniformly
bounded from below by a nonzero constant, the necessary and sufficient condition for
$J(\mathbf{x})$ to be finite is
\begin{equation}\label{J-X-finite}
\int_y^{\infty}L(x)xdx<\infty
\end{equation}
for a sufficiently large $y$~\cite{Da71}.

To investigate percolation-based connectivity of directed SINR graphs, we make the
following assumptions on the signal attenuation function $L(\cdot)$:
\begin{itemize}
\item[(i)] $L(x)< 1, \forall x\in (0, \infty)$; \item[(ii)] $L(0)>\frac{\beta
N_0}{p_{min}}$; and \item[(iii)] $L(x)$ is continuous and strictly decreasing in $x$.
\end{itemize}
The first assumption reflects the fact that the signal power cannot be amplified by
transmitting over a wireless channel. With conditions (i)-(iii), \eqref{J-X-finite} is
guaranteed. Although the last two assumptions are introduced for technical convenience,
they are also practical in real wireless networks.

\section{Percolation in Wireless Networks without Interference}

For interference-free wireless networks, $\gamma=0$. To simplify notation, we let
$\overrightarrow{G}(\mathcal{X}_n,\mathcal{P})$, $G(\mathcal{X}_n,\mathcal{P})$ and
$G'(\mathcal{X}_n,\mathcal{P})$ denote $\overrightarrow{G}(\mathcal{X}_n,\mathcal{P},0)$,
$G(\mathcal{X}_n,\mathcal{P},0)$ and $G'(\mathcal{X}_n,\mathcal{P},0)$, respectively.

When $\gamma=0$, the SINR \eqref{beta-ij} becomes
\begin{equation}\label{eq:beta-i-j-nointerference}
\beta_{ij}=\frac{P_iL(d_{ij})}{N_0}.
\end{equation}
Thus, in $\overrightarrow{G}(\mathcal{X}_n,\mathcal{P})$, there is a directed link from
node $i$ to node $j$ if $L(d_{ij})\geq \frac{N_0\beta}{P_i}$. Since $L(\cdot)$ is
strictly decreasing, this condition becomes $d_{ij}\leq
L^{-1}\left(\frac{N_0\beta}{P_i}\right)$. Let
\begin{equation}\label{eq:R_i}
R_i=L^{-1}\left(\frac{N_0\beta}{P_i}\right),
\end{equation}
where $\underline{r}\leq R_i \leq \bar{r}$ and
\begin{equation}\label{eq:r-underline}
\underline{r}=L^{-1}\left(\frac{N_0\beta}{p_{min}}\right)\end{equation} and
\begin{equation}\label{eq:R_i}
\bar{r}=L^{-1}\left(\frac{N_0\beta}{p_{max}}\right).
\end{equation}
Then,
$\overrightarrow{G}({\cal X}_n, \mathcal{P})$ is the ensemble of graphs where a directed
link exists from node $i$ to node $j$ if $d_{ij}\leq R_i$, $G({\cal X}_n, \mathcal{P})$
is the ensemble of graphs where an undirected link exists between nodes $i$ and $j$
whenever $d_{ij}\leq \min\{R_i,R_j\}$, and $G'({\cal X}_n, \mathcal{P})$ is the ensemble
of graphs where nodes $i$ and $j$ are connected by an undirected link whenever
$d_{ij}\leq \max\{R_i,R_j\}$.

As $n$ and $A$ become large with $n/A=\lambda$ fixed, $G({\cal X}_n, \mathcal{P})$,
$G'({\cal X}_n, \mathcal{P})$ and $\overrightarrow{G}({\cal X}_n, \mathcal{P})$ converge
in distribution to (infinite) graphs $G(\mathcal{H}_{\lambda},\mathcal{P})$,
$G'(\mathcal{H}_{\lambda},\mathcal{P})$ and
$\overrightarrow{G}(\mathcal{H}_{\lambda},\mathcal{P})$ induced by homogeneous Poisson
point processes with density $\lambda>0$, respectively \cite{Pe03, BoRi06}. According to
continuum percolation theory, there exists a positive and finite critical density
$\lambda_c(\mathcal{P})$ ($\lambda'_c(\mathcal{P})$) for
$G(\mathcal{H}_{\lambda},\mathcal{P})$ ($G'(\mathcal{H}_{\lambda},\mathcal{P})$) such
that when $\lambda>\lambda_c(\mathcal{P})$ ($\lambda>\lambda'_c(\mathcal{P})$), there is
a unique component that contains $\Theta(n)$ nodes\footnote{We say $f(n) = O(g(n))$ if
there exists $n_0 > 0$ and constant $c_0$ such that $f(n) \leq c_0g(n)~\forall n \geq
n_0$. We say $f(n) = \Omega(g(n))$ if $g(n) = O(f(n))$. Finally, we say $f(n) =
\Theta(g(n))$ if $f(n) = O(g(n))$ and $f(n) = \Omega(g(n))$.} of
$G(\mathcal{X}_n,\mathcal{P})$ ($G'(\mathcal{X}_n,\mathcal{P})$) a.a.s.\footnote{ An
event is said to be asymptotic almost sure (abbreviated a.a.s.) if it occurs with a
probability converging to 1 as $n \rightarrow \infty$.} This largest component is called
the \emph{giant component}. When $\lambda<\lambda_c(\mathcal{P})$
($\lambda<\lambda'_c(\mathcal{P})$), there is no giant component a.a.s.~\cite{MeRo96,
Pe03, BoRi06}

A useful observation is that $(i,j)\in G(\mathcal{H}_{\lambda},\mathcal{P})$ implies
$(i,j)\in \overrightarrow{G}(\mathcal{H}_{\lambda},\mathcal{P})$ and $(j,i)\in
\overrightarrow{G}(\mathcal{H}_{\lambda},\mathcal{P})$. Also, $(i,j)\in
\overrightarrow{G}(\mathcal{H}_{\lambda},\mathcal{P})$ implies $(i,j)\in
G'(\mathcal{H}_{\lambda},\mathcal{P})$. This relationship will be useful in the following
analysis.

\subsection{Critical Phenomenon}

We now define four types of connected components with respect to a node in
$\overrightarrow{G}({\cal X}_n, \mathcal{P})$. For $\overrightarrow{G}({\cal X}_n,
\mathcal{P})$, we use the notation $u\rightarrow v$ to mean that there is a directed path
from $u$ to $v$. Similarly, we use the notation $u\leftrightarrow v$ to mean that there
is a directed path from $u$ to $v$ and one from $v$ to $u$ as well.

\vspace{0.1in}%
\begin{definition} For a node $u\in \overrightarrow{G}(\mathcal{H}_{\lambda},\mathcal{P})$, the
\emph{in-component} $W_{in}(u)$ is the set of nodes which can reach node $u$, i.e.,
\begin{equation}\label{eq:W-in}
W_{in}(u)\triangleq\{v: v\in \overrightarrow{G}(\mathcal{H}_{\lambda},\mathcal{P}),
v\rightarrow u\}.
\end{equation}
The \emph{out-component} $W_{out}(u)$ is the set of nodes which can be reached from node
$u$, i.e.,
\begin{equation}\label{eq:W-out}
W_{out}(u)\triangleq\{v: v\in \overrightarrow{G}(\mathcal{H}_{\lambda},\mathcal{P}),
u\rightarrow v\}.
\end{equation}
The \emph{weakly connected component} $W_{weak}(u)$ is the set of nodes that are either
in the in-component or the out-component of node $u$, i.e.,
\begin{equation}\label{eq:W-weak}
W_{weak}(u)\triangleq W_{in}(u)\cup W_{out}(u).
\end{equation}
The \emph{strongly connected component} $W_{strong}(u)$ is the set of nodes that are in
both the in-component and the out-component of node $u$, i.e.,
\begin{equation}\label{eq:W-strong}
W_{strong}(u)\triangleq W_{in}(u)\cap W_{out}(u).
\end{equation}
\end{definition}
\vspace{0.1in}%

Corresponding to these four types of components, there are four types of phase
transitions. For instance, a directed graph is said to be in the in-component
supercritical phase if with probability 1 there exists an infinite in-component, and
in-component subcritical phase otherwise. In the next subsection we will investigate the
existence of such a phase transition in directed random geometric graphs. Before that, we
define the critical densities. Formally, let
$\mathcal{H}_{\lambda,\mathbf{0}}=\mathcal{H}_{\lambda}\cup \{\mathbf{0}\}$, i.e., the
union of the origin and the infinite homogeneous Poisson point process with density
$\lambda$. Note that in a random geometric graph induced by a homogeneous Poisson point
process, the choice of the origin can be arbitrary.

\vspace{0.1in}%
\begin{definition} Let $p_{\infty}^{in}(\lambda)$, $p_{\infty}^{out}(\lambda)$, $p_{\infty}^{weak}(\lambda)$
and $p_{\infty}^{strong}(\lambda)$ be the probabilities that the in-component,
out-component, weakly connected component and strongly connected component containing the
origin has an infinite number of nodes of the graph $G(\mathcal{H}_{\lambda,\mathbf{0}},
\mathcal{P})$, respectively. The critical densities for the in-component, out-component,
weakly connected component and strongly connected component phase transitions are defined
respectively as
\begin{eqnarray}
\lambda_{in} & \triangleq & \inf\{\lambda:
p_{\infty}^{in}(\lambda)>0\},\label{lambda-in}\\
\lambda_{out} & \triangleq & \inf\{\lambda:
p_{\infty}^{out}(\lambda)>0\},\label{lambda-out}\\
\lambda_{weak} & \triangleq & \inf\{\lambda:
p_{\infty}^{weak}(\lambda)>0\},\label{lambda-weak}\\
\lambda_{strong} & \triangleq & \inf\{\lambda:
p_{\infty}^{strong}(\lambda)>0\}.\label{lambda-strong}
\end{eqnarray}
\end{definition}
\vspace{0.1in}%

By the same argument as in the proof for Theorem 9.19 in~\cite{Pe03}, it can be shown
that when $p_{\infty}^{in}(\lambda)>0$, $G(\mathcal{H}_{\lambda,\mathbf{0}},
\mathcal{P})$ has precisely one infinite in-component with probability 1. It can also be
shown that this infinite in-component contains a constant fraction of nodes in the
network a.a.s.~\cite{Pe03}. The same results hold for the out-component, weakly connected
component and strongly connected component phase transitions.

Theorem \ref{Theorem-Critical-Densities-Equalities} below asserts that all four critical
densities of $\overrightarrow{G}(\mathcal{H}_{\lambda},\mathcal{P})$ are actually equal,
i.e., all four types of phase transitions occur at the same time. The result is not
entirely intuitive since it is plausible to imagine that a directed random geometric
graph experiences three or four steps of phase transitions. That is, as the density
increases, first an infinite weakly connected component emerges, then an infinite
in-component and an infinite out-component appear (successively or instantaneously), and
finally an infinite strongly connected component forms. Nonetheless, we will see that
these four types of infinite components form at exactly the same time.

\vspace{0.1in}%
\begin{theorem}\label{Theorem-Critical-Densities-Equalities}
$\lambda_{in}=\lambda_{out}=\lambda_{weak}=\lambda_{strong}$.
\end{theorem}
\vspace{0.1in}%

\emph{Proof:} Since the choice of the origin in
$\overrightarrow{G}(\mathcal{H}_{\lambda},\mathcal{P})$ can be arbitrary, when
$\lambda>\lambda_{in}$, for any node $u\in
\overrightarrow{G}(\mathcal{H}_{\lambda},\mathcal{P})$,
$\mbox{Pr}\{|W_{in}(u)|=\infty\}>0$. Because all nodes in
$\overrightarrow{G}(\mathcal{H}_{\lambda},\mathcal{P})$ are distributed according to a
homogeneous Poisson process, for any node
$u\in\overrightarrow{G}(\mathcal{H}_{\lambda},\mathcal{P})$, the probability
$\mbox{Pr}\{v\in W_{in}(u)\}$ is identical for all
$v\in\overrightarrow{G}(\mathcal{H}_{\lambda},\mathcal{P})$, we have $\mbox{Pr}\{v\in
W_{in}(u)\}>0$ for all $v\in
\overrightarrow{G}(\mathcal{H}_{\lambda},\mathcal{P})$.\footnote{Note that, $u$ and $v$
are nodes with fixed labels and random positions.} Equivalently, for any node $v\in
\overrightarrow{G}(\mathcal{H}_{\lambda},\mathcal{P})$, $\mbox{Pr}\{u\in W_{out}(v)\}>0$
for all $u\in \overrightarrow{G}(\mathcal{H}_{\lambda},\mathcal{P})$, which implies that
$\mbox{Pr}\{|W_{out}(v)|=\infty\}>0$. That is $\lambda>\lambda_{out}$. Similarly, we can
show when $\lambda<\lambda_{in}$, then $\lambda<\lambda_{out}$, and therefore
$\lambda_{in}=\lambda_{out}$.

It is obvious that $\lambda_{weak}\leq\lambda_{in}=\lambda_{out}$. On the other hand, if
$\lambda>\lambda_{weak}$, then either $p_{\infty}^{in}(\lambda)>0$ or
$p_{\infty}^{out}(\lambda)>0$. Hence we have $\lambda_{weak}=\lambda_{in}=\lambda_{out}$.

It is also obvious that $\lambda_{strong}\geq\lambda_{in}=\lambda_{out}$. Since the
events $\{|W_{in}(\mathbf{0})|=\infty\}$ and $\{|W_{out}(\mathbf{0})|=\infty\}$ are
increasing events,\footnote{An event $A$ is called increasing if $I_A(G)\leq I_A(G')$
whenever graph $G$ is a subgraph of $G'$, where $I_A$ is the indicator function of $A$.
An event $A$ is called decreasing if $A^{c}$ is increasing. For details, please
see~\cite{Gr99, MeRo96, Pe03}.} by the FKG inequality~\cite{Gr99, MeRo96, Pe03}, we have
$\mbox{Pr}\{\{|W_{in}(\mathbf{0})|=\infty\}\cap\{|W_{out}(\mathbf{0})|=\infty\}\}\geq
\mbox{Pr}\{|W_{in}(\mathbf{0})|=\infty\} \mbox{Pr}\{|W_{out}(\mathbf{0})|=\infty\}$,
i.e., $p_{\infty}^{strong}(\lambda)\geq
p_{\infty}^{in}(\lambda)p_{\infty}^{out}(\lambda)$. Hence, if
$\lambda>\lambda_{in}=\lambda_{out}$, the graph is in the strongly connected component
supercritical phase. Thus we have $\lambda_{strong}\leq\lambda_{in}=\lambda_{out}$, and
therefore $\lambda_{strong}=\lambda_{in}=\lambda_{out}$. \qed

Since all four critical densities are equal, we now define
$\overrightarrow{\lambda}_c(\mathcal{P})$ as the critical density at which all four types
of phase transitions take place in
$\overrightarrow{G}(\mathcal{H}_{\lambda},\mathcal{P})$, i.e.,
$\overrightarrow{\lambda}_c(\mathcal{P})
=\lambda_{in}=\lambda_{out}=\lambda_{weak}=\lambda_{strong}$.

\subsection{Bounds for the Critical Densities}

Instead of directly showing that there exists a positive and finite critical density
$\overrightarrow{\lambda}_c(\mathcal{P})$ for
$\overrightarrow{G}(\mathcal{H}_{\lambda},\mathcal{P})$, we provide tight analytical
upper and lower bounds on $\overrightarrow{\lambda}_c(\mathcal{P})$. To accomplish this,
we couple $\overrightarrow{G}(\mathcal{H}_{\lambda},\mathcal{P})$ with
$G(\mathcal{H}_{\lambda},\mathcal{P})$ and $G'(\mathcal{H}_{\lambda},\mathcal{P})$. Due
to the relationship between $G(\mathcal{H}_{\lambda},\mathcal{P})$,
$\overrightarrow{G}(\mathcal{H}_{\lambda},\mathcal{P})$ and
$G'(\mathcal{H}_{\lambda},\mathcal{P})$, it is easy to see that
$\lambda_c'(\mathcal{P})\leq \overrightarrow{\lambda}_c(\mathcal{P})\leq
\lambda_c(\mathcal{P})$. By employing the cluster coefficient method~\cite{KoYe07-1,
KoYe07-2} and the re-normalization method~\cite{MeRo96}, we proved a lower bound on
$\lambda_c'(\mathcal{P})$ and an upper bound on $\lambda_c(\mathcal{P})$, respectively.

In~\cite{KoYe07-1, KoYe07-2}, a new method has been proposed to provide lower bounds on
the critical densities for $d$-dimensional random geometric graphs. The methodology is
based on the clustering effect in random geometric graphs which can be characterized by
$t$-th order cluster coefficients. In the same manner, we define the cluster coefficients
for $G'(\mathcal{H}_{\lambda}, \mathcal{P})$ as follows.

\vspace{0.1in}%
\begin{definition} For any integer $t\geq 3$, suppose $v_1, \ldots,
 v_{t-1} \in G'(\mathcal{H}_{\lambda}, \mathcal{P})$
form a single chain, i.e., they satisfy the following properties:
\begin{itemize}
\item[i)] For each $j=1,2,...,t-2$, $(v_j,v_{j+1})\in E$. \item[ii)]  For all $1\leq
j,k\leq t-1$, $(v_j,v_k)\notin E$ for $|j-k| > 1$,
\end{itemize}
where $E$ denotes the set of links in $G'(\mathcal{H}_{\lambda}, \mathcal{P})$. Then the
\emph{$t$-th order cluster coefficient $C'_t$} is defined to be the conditional
probability that a node $v_t$ is adjacent to at least one of the nodes $v_2,...,v_{t-1},$
given that $v_t$ is adjacent to $v_1$ (averaging over all the possible positions ${\mb
X_{v_2}},\ldots,{\mb X_{v_{t-1}}}$ in $\mathbb{R}^2$ of the points $v_2$, $\ldots,
v_{t-1}$ satisfying conditions (i) and (ii)).
\end{definition}
\vspace{0.1in}%

In general, evaluating $C'_t$, $t\geq 4$ for $G'(\mathcal{H}_{\lambda},\mathcal{P})$ is
quite difficult. Fortunately, the cluster coefficient $C_3'$ can be obtained through a
geometrical calculation. The details are given in Appendix A.

\vspace{0.1in}%
\begin{theorem}\label{Theorem-Bound}
For 2-dimensional directed SINR graphs, we have for any integer $t\geq 3$,
\begin{equation}\label{lambda-lower-bound}
\overrightarrow{\lambda}_c(\mathcal{P}) \geq
\frac{1}{[1-C_t']\pi\int_{\underline{r}}^{\bar{r}}g(r)f_R(r)dr},
\end{equation}
where $C_t'$ is the $t$-th cluster coefficient for
$G'(\mathcal{H}_{\lambda},\mathcal{P})$,
\begin{eqnarray}
f_R(r) & = &
-f_P\left(\frac{N_0\beta}{L(r)}\right)\frac{N_0\beta}{L(r)^2}\frac{dL(r)}{dr},\label{eq:f-R-r}\\
g(r) & = & r^2 + 2\int_r^{\bar{r}} r'[1-F_R(r')]dr',\label{eq:g-r}
\end{eqnarray}
and
\begin{equation}\label{lambda-upper-bound}
\overrightarrow{\lambda}_c(\mathcal{P}) < \frac{4\ln
2}{\left(\pi-6\varphi-\frac{3\sqrt{3}(\sqrt{5}-1)}{8}\right)\underline{r}^2},
\end{equation}
where $\varphi=\sin^{-1}\left(\frac{1}{4}\right)$ and
$\underline{r}=L^{-1}\left(\frac{N_0\beta}{p_{min}}\right)$.
\end{theorem}
\vspace{0.1in}%

\emph{Proof:} To show the lower bound on $\overrightarrow{\lambda}_c(\mathcal{P})$, we
employ the cluster coefficient method~\cite{KoYe07-1, KoYe07-2} to obtain lower bounds on
$\lambda_c'(\mathcal{P})$. Consider a node $i \in G'(\mathcal{H}_{\lambda},\mathcal{P})$
at a given location ${\mb x}_i$ with given radius $r$. All the other nodes lying within
$\mathcal{A}({\mb x_i},r)$---the circular region centered at ${\mb x_i}$ with radius $r$,
are adjacent to node $i$. The expected number of nodes in $\mathcal{A}({\mb x_i},r)$ is
$\lambda \pi r^2$.

A node $j$ within $\mathcal{D}({\mb x_i},r,\bar{r})$---the annulus centered at ${\mb
x_i}$ with inner radius of $r$ and outer radius of $\bar{r}$, is adjacent to $i$ if and
only if $R_j\geq||\mathbf{x}_j-\mathbf{x}_i||$. The expected number of nodes satisfying
this condition is
\[
\int_r^{\bar{r}} \lambda 2\pi r'\int_{r'}^{\bar{r}} f_R(r'')dr''dr'.
\]
Thus the mean degree for node $i$ is
\[
\mu(r)= \lambda \pi r^2 + \!\int_r^{\bar{r}}\lambda 2\pi r'
\int_{r'}^{\bar{r}}f_R(r'')dr''dr' =\lambda \pi g(r).
\]
The mean degree of $G'(\mathcal{H}_{\lambda},\mathcal{P})$ is
\[
\mu' = \lambda \pi\int_{\underline{r}}^{\bar{r}}g(r)f_R(r)dr.
\]
Apply the results of
Theorem 1 in~\cite{KoYe07-1, KoYe07-2} to $G'(\mathcal{H}_{\lambda},\mathcal{P})$, we
have $\mu'_c(\mathcal{P})\geq\frac{1}{1-C_t'}$. Thus
\[
\lambda_c'(\mathcal{P})\geq \frac{1}{(1-C_t')\pi\int_{\underline{r}}^{\bar{r}}g(r)f_R(r)dr},
\]
which yields the lower bound for $\overrightarrow{\lambda}_c(\mathcal{P})$. Because
$R=L^{-1}\left(\frac{N_0\beta}{P}\right)$, we have~\eqref{eq:f-R-r}.

\begin{figure}[t!]
\centering
\includegraphics[width=2.6in]{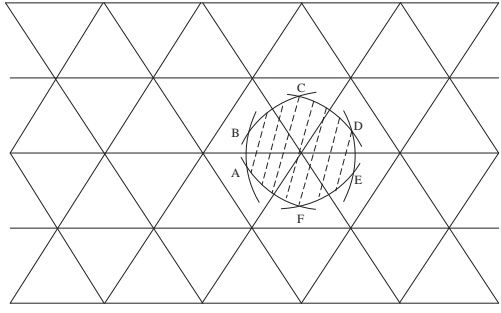}
\caption{Triangular lattice with flower shaped region around every
site.}\label{fig:TriangleFlower}
\end{figure}

To prove the upper bound, we use a mapping between the continuum percolation model and a
discrete site percolation model on a triangular lattice to obtain an upper bound on
$\lambda_c(\mathcal{P})$ and therefore an upper bound on
$\overrightarrow{\lambda}_c(\mathcal{P})$. Similar methods have been used
in~\cite{MeRo96}. Let $\mathcal{L}_T$ be the triangular lattice with edge length
$\underline{r}/2$. Each site is enclosed by a flower shaped region, which is formed by
six arcs of circles. Each of the circles has radius of length $\underline{r}/2$ and is
centered at the midpoint of each edge adjacent to the site. This is shown in
Figure~\ref{fig:TriangleFlower}.

We say a site is open if there is at least one node of
$G(\mathcal{H}_{\lambda},\mathcal{P})$ in the corresponding flower shaped region and
closed otherwise. If any two adjacent sites are both open, then the flower shaped regions
corresponding to these two sites both contain at least one node of
$G(\mathcal{H}_{\lambda},\mathcal{P})$. Because $R$ is lower bounded by $\underline{r}$,
the nodes within these flower shaped regions must be directly connected by a link.
Consequently, if site percolation on the triangular lattice occurs, i.e., there is an
infinite cluster of adjacent open sites, then percolation occurs in the continuum model.
Since the underlying point process for $G(\mathcal{H}_{\lambda},\mathcal{P})$ is a
homogeneous Poisson process and flower shaped regions of different sites are disjoint,
the probability $p_o$ that each site is open is identical and the events are independent
of each other. Furthermore,
\[
p_o=1-e^{-\lambda S_f},
\]
where $S_f$ is the area of the flower shaped region, which can
be calculated as
\[
S_f=\frac{1}{4}\left(\pi-6\varphi-\frac{3\sqrt{3}(\sqrt{5}-1)}{8}\right)\underline{r}^2,
\]
where $\varphi=\sin^{-1}\left(\frac{1}{4}\right)$.

From the theory of discrete percolation, we know that for site percolation on triangular
lattices, the critical probability is $p_c=\frac{1}{2}$~\cite{Gr99}. Thus if
$1-e^{-\lambda S_f}>\frac{1}{2}$, percolation occurs in
$G(\mathcal{H}_{\lambda},\mathcal{P})$. Therefore,
\[
\lambda_c(\mathcal{P}) < \frac{4\ln
2}{\left(\pi-6\varphi-\frac{3\sqrt{3}(\sqrt{5}-1)}{8}\right)\underline{r}^2},
\]
and inequality \eqref{lambda-upper-bound} holds.\qed

A special case of the above model is the directed random geometric graph
$\overrightarrow{G}(\mathcal{H}_{\lambda},(a,b))$ with binary distributed transmission
radii, i.e., $\mbox{Pr}\{R=a\}=p_a$ and $\mbox{Pr}\{R=b\}=p_b$, where $p_a\geq 0,p_b\geq
0, p_a+p_b=1$ and $0<a\leq b$. For this model, we calculate the cluster coefficient
(detailed analysis is given in Appendix B) as
\begin{equation}\label{bar-C}
\bar{C} = (p_b^3+p_a^3+3p_b^2p_a)C + p_bp_a^2\left(\frac{2}{\pi b^4}\right)\cdot\int_0^b
\left[(\phi_1+\theta_1)(a^2+b^2)+h\sin \theta_1 (a+b)\right]hdh,
\end{equation}
where $\phi_1 = \cos^{-1}\left(\frac{h^2+a^2-b^2}{2ah}\right)$ and
$\theta_1=\cos^{-1}\left(\frac{h^2+b^2-a^2}{2bh}\right)$.

In light of Theorem \ref{Theorem-Bound}, we have

\vspace{0.1in}%
\begin{corollary}\label{Corollary-Special-Case-Bound}
Let $\overrightarrow{\lambda}_c(a,b)$ be the critical density for
$\overrightarrow{G}(\mathcal{H}_{\lambda},(a,b))$, then,
\begin{equation}\label{lambda-c-special-case-bound}
\frac{1}{\pi(1-\bar{C})b^2}\!\leq\! \overrightarrow{\lambda}_c(a,b)\! <\! \frac{4\ln
2}{\left(\pi-6\varphi-\frac{3\sqrt{3}(\sqrt{5}-1)}{8}\right)a^2},
\end{equation}
where $\varphi=\sin^{-1}\left(\frac{1}{4}\right)$.
\end{corollary}
\vspace{0.1in}%

\section{Percolation in Wireless Networks with Interference}

Now consider the scenario with $\gamma>0$. In this case the transmission region of each
node is irregular instead of circular. Nevertheless, the definitions of the four types of
phase transitions are still applicable. Percolation in
$\overrightarrow{G}(\mathcal{H}_{\lambda},\mathcal{P},\gamma)$ depends not only on
$\lambda$, but also on $\gamma$. As before, we can show that the critical value of
$\gamma$ is the same for all four types of phase transitions. That is, in
$\overrightarrow{G}(\mathcal{H}_{\lambda},\mathcal{P},\gamma)$, if $\gamma$ is strictly
less than the critical value, there exists a unique giant strongly (in, out, weakly)
connected component. Formally, define

\vspace{0.1in}%
\begin{definition}\label{gamma-c}
\vspace{-0.05in}%
\begin{eqnarray*}
\overrightarrow{\gamma}_c(\lambda) & \triangleq & \sup\{\gamma:
\overrightarrow{G}(\mathcal{H}_{\lambda},\mathcal{P},\gamma) \mbox{ is percolated, }
\lambda>\overrightarrow{\lambda}_c\},\\
\gamma_c(\lambda)& \triangleq & \sup\{\gamma: G(\mathcal{H}_{\lambda},\mathcal{P},\gamma) \mbox{
is percolated,
} \lambda>\lambda_c\},\\
\gamma'_c(\lambda)& \triangleq & \sup\{\gamma:
G'(\mathcal{H}_{\lambda},\mathcal{P},\gamma) \mbox{ is percolated, }
\lambda>\lambda'_c\},
\end{eqnarray*}
where $\overrightarrow{\lambda}_c(\mathcal{P})$, $\lambda_c(\mathcal{P})$ and
$\lambda'_c(\mathcal{P})$ are the critical densities for
$\overrightarrow{G}(\mathcal{H}_{\lambda},\mathcal{P})$,
$G(\mathcal{H}_{\lambda},\mathcal{P})$ and $G'(\mathcal{H}_{\lambda},\mathcal{P})$,
respectively.
\end{definition}
\vspace{0.1in}%

The critical values of $\gamma$ depend on the network density $\lambda$, the distribution
of the transmission power $f_P(p)$, background noise power $N_0$, and the threshold
$\beta$. To show the existence of $\overrightarrow{\gamma}_c(\lambda)$ and to provide
lower and upper bounds for it, we employ the same technique as in the previous section.
By coupling $\overrightarrow{G}(\mathcal{H}_{\lambda},\mathcal{P})$ with
$G(\mathcal{H}_{\lambda},\mathcal{P},\gamma)$ and
$G'(\mathcal{H}_{\lambda},\mathcal{P},\gamma)$, it is easy to see that $\gamma_c(\lambda)
\leq \overrightarrow{\gamma}_c(\lambda) \leq \gamma'_c(\lambda)$. Thus, by obtaining
lower bounds on $\gamma_c(\lambda)$ and upper bounds on $\gamma_c'(\lambda)$, we obtain
bounds on $\overrightarrow{\gamma}_c(\lambda)$.

Let $\lambda_c(\underline{r})$ be the critical density for
$G(\mathcal{H}_{\lambda},\underline{r})$ with $\underline{r}=L^{-1}\left(\frac{\beta
N_0}{p_{min}}\right)$. It was shown in~\cite{DoBaTh05, DoFrMaMeTh06} that for any
$\lambda>\lambda_c(\underline{r})$, there exists a constant $c_1>0$ such that
$\gamma_c(\lambda)>\frac{c_1}{\lambda}$. This provides a lower bound for
$\overrightarrow{\gamma}_c(\lambda)$.

To obtain an upper bound for $\overrightarrow{\gamma}_c(\lambda)$, we employ a new
mapping technique to obtain an upper bound on $\gamma'_c(\lambda)$.

\vspace{0.1in}%
\begin{theorem}\label{Theorem-gamma-c-bound-2} For any $\lambda\geq \lambda_c'(\mathcal{P})$,
there exist constants $0<c_2<\infty$ and $0<c_2'<\lambda$, such that
\begin{equation}\label{eq:gamma-c-bound}
\overrightarrow{\gamma}_c(\lambda)\leq\frac{c_2}{\lambda-c_2'}.
\end{equation}
\end{theorem}

\begin{figure}[t!]
\centering
\includegraphics[width=2.8in]{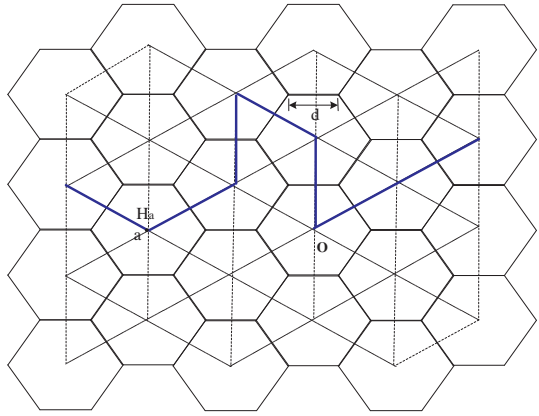}
\caption{Paths starting from the origin in the dual triangular lattice $\mathcal{L}'_T$
and hexagonal lattice $\mathcal{L}_H$}\label{fig:HexigonalLatticePath}
\end{figure}

\emph{Proof:} To prove \eqref{eq:gamma-c-bound}, we show that $\gamma'_c(\lambda)\leq
\frac{c_2}{\lambda-c_2'}$. Map $G'(\mathcal{H}_\lambda,\mathcal{P},\gamma)$ to a
hexagonal lattice $\mathcal{L}_H$ with edge length $d > L^{-1}\left(\frac{\beta
N_0}{p_{max}}\right)$ as shown in Figure~\ref{fig:HexigonalLatticePath}. Then,
$L(d)<\frac{\beta N_0}{p_{max}}$. Therefore, there is no link between any nodes within
two different hexagons that do not share common edges.

Let the dual triangular lattice be $\mathcal{L}'_T$\footnote{The construction of
$\mathcal{L}'_T$ is as follows: let each vertex of $\mathcal{L}'_T$ be located at the
center of a hexagon of $\mathcal{L}_H$.}, and the hexagon of $\mathcal{L}_H$ centered at
vertex $a$ of $\mathcal{L}_T'$ be $H_a$ (see Figure~\ref{fig:HexigonalLatticePath}).
Denote by $N(H_a)$ the number of nodes of $G'(\mathcal{H}_\lambda,\mathcal{P},\gamma)$
contained in $H_a$, then $N(H_a)$ has a Poisson distribution with mean
\[
E[N(H_a)]=\lambda\frac{3\sqrt{3}}{2}d^2.
\]
Now note that if the number of nodes contained
in a hexagon is strictly greater than
\begin{equation}\label{eq:N-prime}
N' \triangleq \left\lceil\frac{p_{max}-\beta N_0}{\gamma\beta p_{min} L(2d)}\right\rceil+2,
\end{equation}
then all the nodes in such a hexagon are isolated (i.e, no two nodes share a link). To
see this note that $0<L(x)<1$ and $L(x)$ is strictly decreasing in $x$. If $N(H_a)> N'$,
then for any nodes $i$ and $j$ in hexagon $H_a$,
\begin{eqnarray*}
\beta_{ij} & = & \frac{P_iL(||\mathbf{X}_i-\mathbf{X}_j||)}{N_0+\gamma\sum_{k\neq
i,j}P_kL(||\mathbf{X}_k-\mathbf{X}_j||)}\\
& \leq & \frac{p_{max}}{N_0+\gamma(N-2)p_{min}L(2d)}\\
& < & \frac{p_{max}}{N_0+\gamma(N'-2)p_{min}L(2d)} \\
& \leq &\beta,
\end{eqnarray*}
and similarly $\beta_{ji}< \beta$. This implies that there is no link between nodes $i$
and $j$. Hence all the nodes in the hexagon $a$ are isolated.

For each vertex $a$ of $\mathcal{L}_T'$ and hexagon $H_a$ of $\mathcal{L}_H$, let $C_a$
be the event that $\{N(H_a)\leq N'\}$. We call hexagon $H_a$ and vertex $a$ \emph{open}
if $C_a$ occurs, and let
\begin{equation}\label{eq:p-o}p_o\triangleq\mbox{Pr}\{C_a\}=\mbox{Pr}\{N(H_a)\leq N'\}.
\end{equation}
Let
\begin{equation}\label{eq:epsilon}
\theta=\frac{\sqrt{10}}{d\sqrt[4]{27}\sqrt{\lambda_c'(\mathcal{P})}},
\end{equation}
then
\[
\theta^2E[N(H_a)]=\theta^2\lambda\frac{3\sqrt{3}}{2}d^2=\frac{5\lambda}{\lambda_c'(\mathcal{P})}.
\]
Now choose $d$ sufficiently large so that $(1-\theta)E[N(H_a)]-3>0$. Let
\begin{equation}\label{gamma-upper-bound}
\gamma_2 = \frac{p_{max}-\beta N_0}{\beta p_{min} L(2d) [(1-\theta)E[N(H_a)]-3]}.
\end{equation}
Since $p_{max}>\beta N_0$ and $(1-\theta)E[N(H_a)]-3>0$, $\gamma_2>0$. Moreover, as
$(1-\theta)E[N(H_a)]\geq N'$, by the Chebyshev's inequality, when $\gamma>\gamma_2$,
\begin{eqnarray*}
p_o & = & \mbox{Pr}\{N(H_a)\leq N'\}\\
& \leq & \mbox{Pr}\{N(H_a)\leq (1-\theta)E[N(H_a)] \}\\
& \leq & \frac{Var[N(H_a)]}{\theta^2 E[N(H_a)]^2}\\
& = & \frac{1}{\theta^2 E[N(H_a)]}\nonumber\\
& = & \frac{\lambda_c'(\mathcal{P})}{5\lambda}\\
& < & \frac{1}{5},
\end{eqnarray*}
where the last inequality is due to $\lambda>\lambda'_c(\mathcal{P})$.

\begin{figure}[t!]
\centering
\includegraphics[width=2.8in]{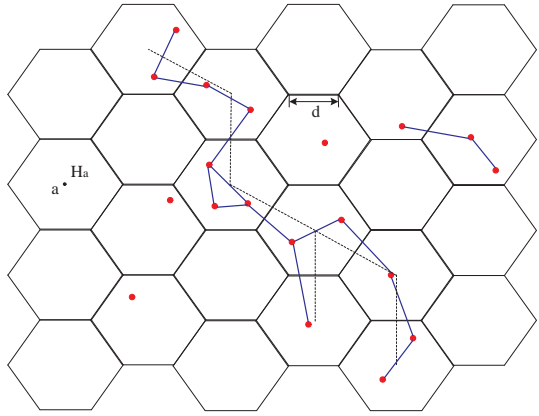}
\caption{An infinite component in $G'(\mathcal{H}_\lambda,\mathcal{P},\gamma)$ implies a
path passing through an infinite number of open vertices in $\mathcal{L}'_T$ (open
hexagons in $\mathcal{L}'_H$).}\label{fig:HexigonalPercolation}
\end{figure}

Note that if there is an infinite component in
$G'(\mathcal{H}_\lambda,\mathcal{P},\gamma)$, there must exist a path passing through an
infinite number of open vertices in $\mathcal{L}'_T$ (open hexagons in $\mathcal{L}_H$),
as illustrated in Figure~\ref{fig:HexigonalPercolation}. This is because along the
infinite component in $G'(\mathcal{H}_\lambda,\mathcal{P},\gamma)$, each hexagon of
$\mathcal{L}_H$ contains at least one node of
$G'(\mathcal{H}_\lambda,\mathcal{P},\gamma)$.

Now choose a path in $\mathcal{L}'_T$ starting from the origin having length $m$. Because
the status (i.e., open or closed) of each vertex $a$ depends only on the number of nodes
contained in the hexagon $H_a$, for different vertices $a$ and $b$, events $C_a$ and
$C_b$ are independent. Thus
\[
\mbox{Pr}\{\exists \mathcal{O}_p(m)\}\leq \xi(m)p_o^{m},
\]
where $\mathcal{O}_p(m)$ is a open path in $\mathcal{L}_T$ starting from the origin with
length $m$, and $\xi(m)$ is the number of such paths. For a path in $\mathcal{L}'_T$ from
the origin, the first edge has six choices for its direction, and all the other edges
have five choices for their directions as illustrated in
Figure~\ref{fig:HexigonalLatticePath}. Therefore, we have
\[
\xi(m)\leq 6\cdot 5^{m-1}.
\]
Consequently,
\begin{equation}\label{eq:open-path-probability}
\mbox{Pr}\{\exists
\mathcal{O}_p(m)\}\leq 6\cdot5^{m-1}p_o^{m}=\frac{6}{5}(5p_o)^m.
\end{equation}
When
$\gamma>\gamma_2$, $p_o<\frac{1}{5}$ and hence $\frac{6}{5}(5p_o)^m$ converges to 0 as $m
\rightarrow \infty$. This implies that there is no infinite path in $\mathcal{L}'_T$
a.a.s., and therefore there is no infinite component in
$G'(\mathcal{H}_\lambda,\mathcal{P},\gamma)$ a.a.s. either. By setting
\begin{equation}\label{c-2}
c_2 = \frac{2\sqrt{3}(p_{max}-\beta N_0)}{9(1-\theta)\beta p_{min} L(2d)d^2}
\end{equation}
and
\begin{equation}\label{c-2-prime}
c'_2 = \frac{2\sqrt{3}}{3(1-\theta)d^2},
\end{equation}
we complete our proof.\qed

From this theorem, we obtain the following corollary which shows that
$\overrightarrow{\gamma_c}(\lambda)$ and $\gamma_c(\lambda)$ have the same asymptotic
behavior with respect to $\lambda$~\cite{DoBaTh05, DoFrMaMeTh06}.
\vspace{0.1in}%
\begin{corollary}\label{Corollary-gamma-c-asymptotic}
For sufficiently large $\lambda$, $\overrightarrow{\gamma}_c(\lambda)=\Theta\left(\frac{1}{\lambda}\right)$.
\end{corollary}

\section{Simulation Studies}

\begin{figure}[t!]
\centerline{ \subfigure[subcritical]{
\includegraphics[width=3.0in]{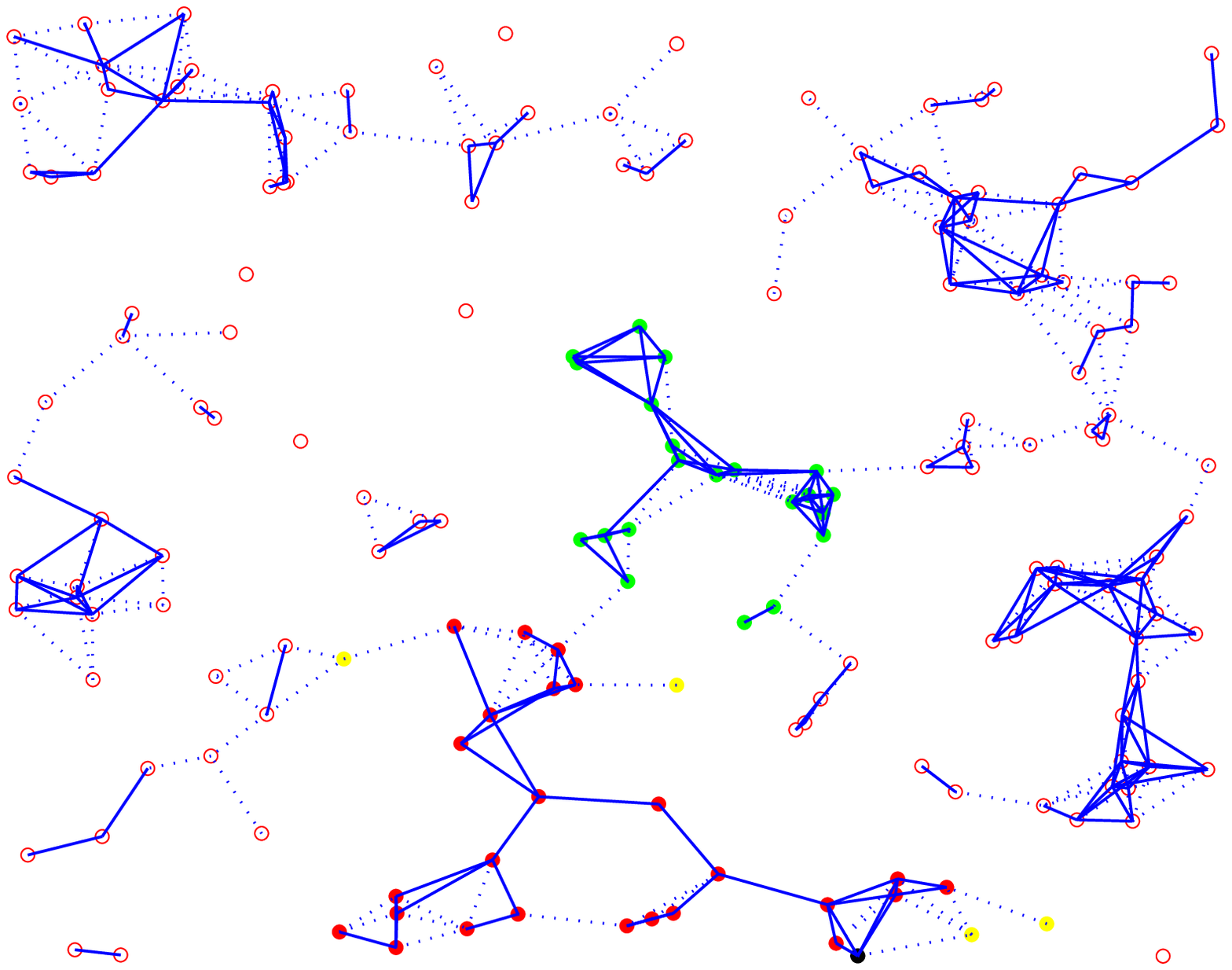}}\hfil
\subfigure[supercritical]{
\includegraphics[width=3.0in]{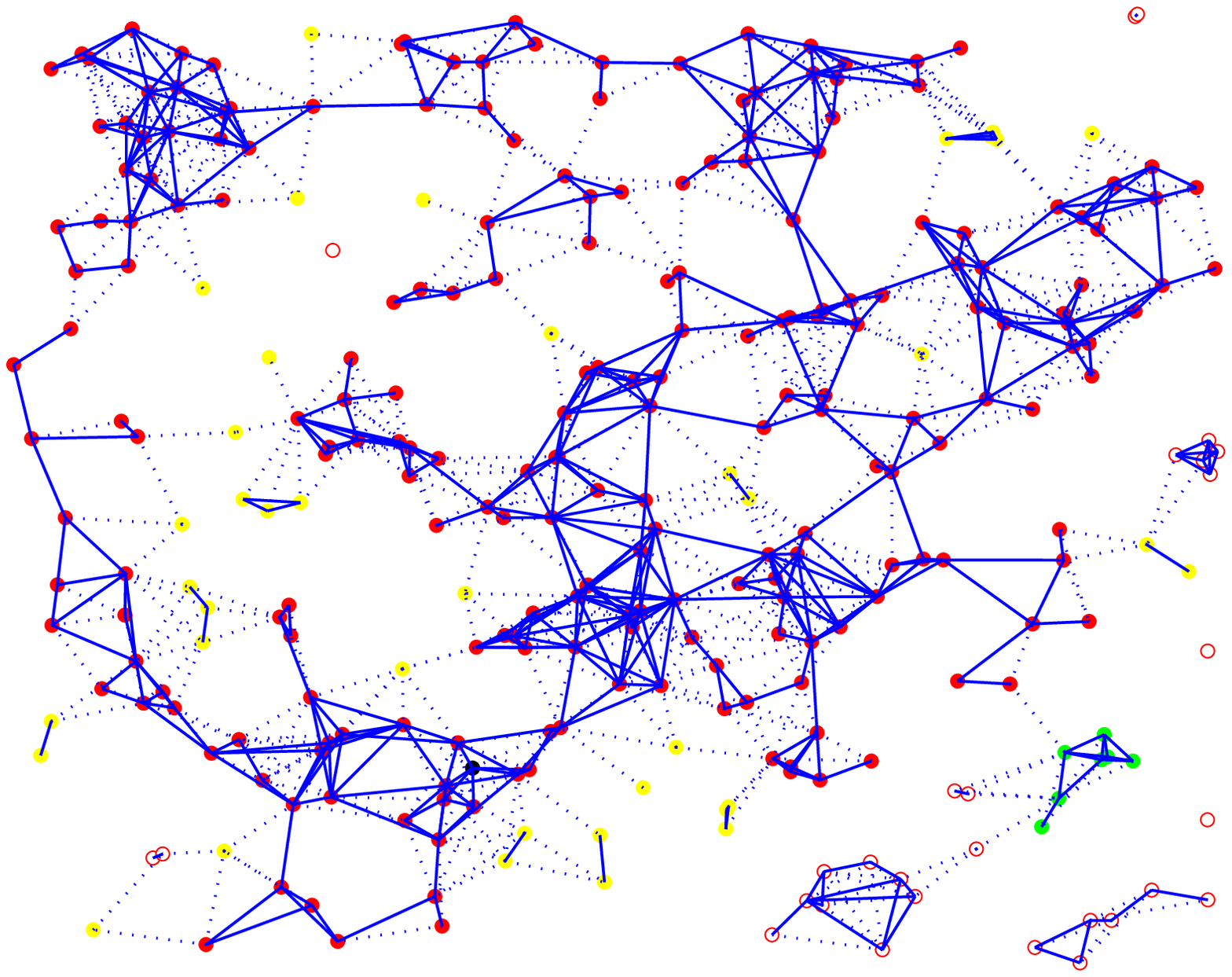}}}
\caption{Percolation in directed SINR graphs without interference where transmission
powers have a binary distribution such that $\mbox{Pr}\{R_i=1\}=0.5$ and
$\mbox{Pr}\{R_i=2\}=0.5$: (a) $\lambda=0.45$, (b) $\lambda=0.75$.}\label{fig:BDRGG1}
\end{figure}

\begin{figure}[t!]
\centerline{ \subfigure[subcritical]{
\includegraphics[width=3.0in]{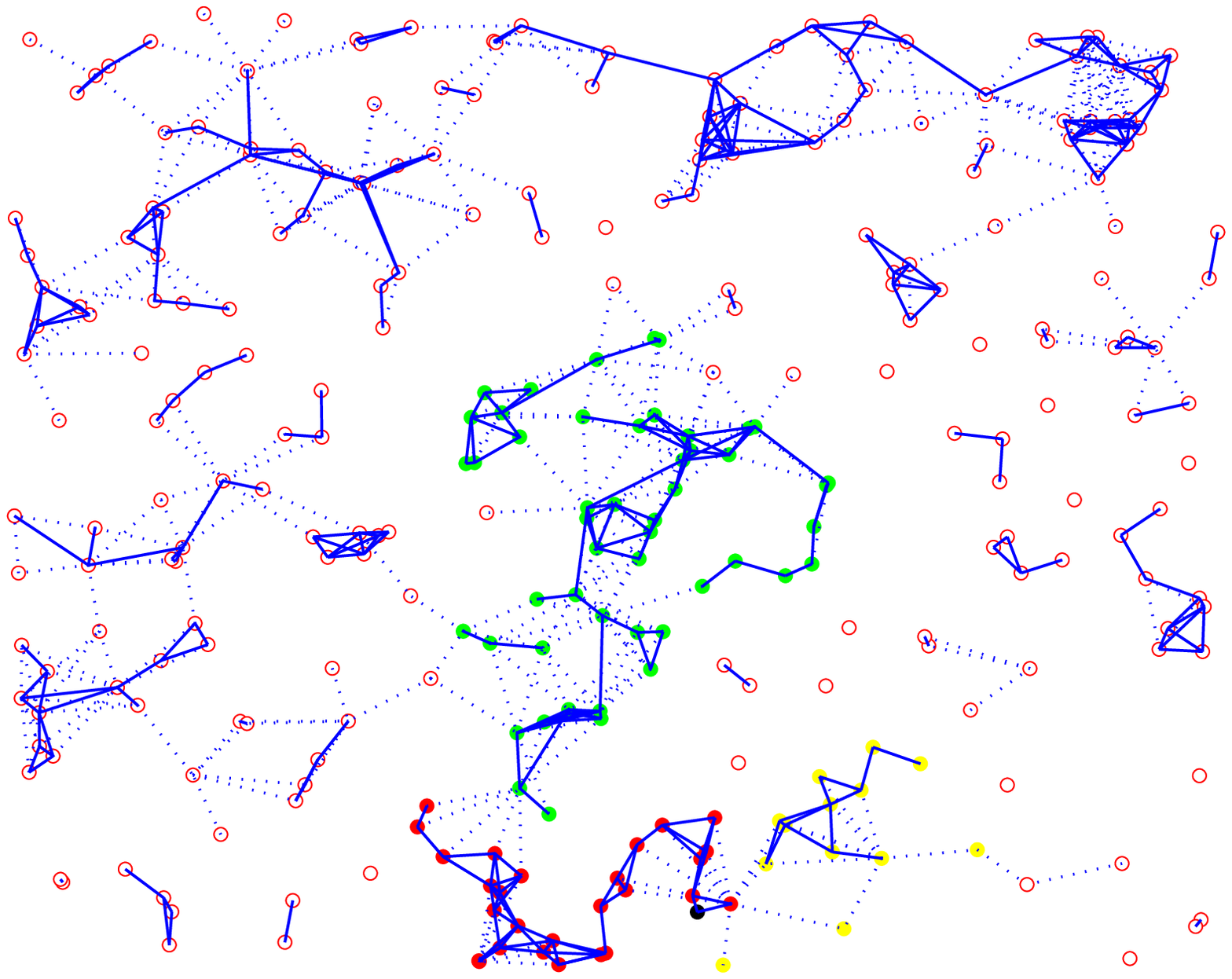}}\hfil
\subfigure[supercritical]{
\includegraphics[width=3.0in]{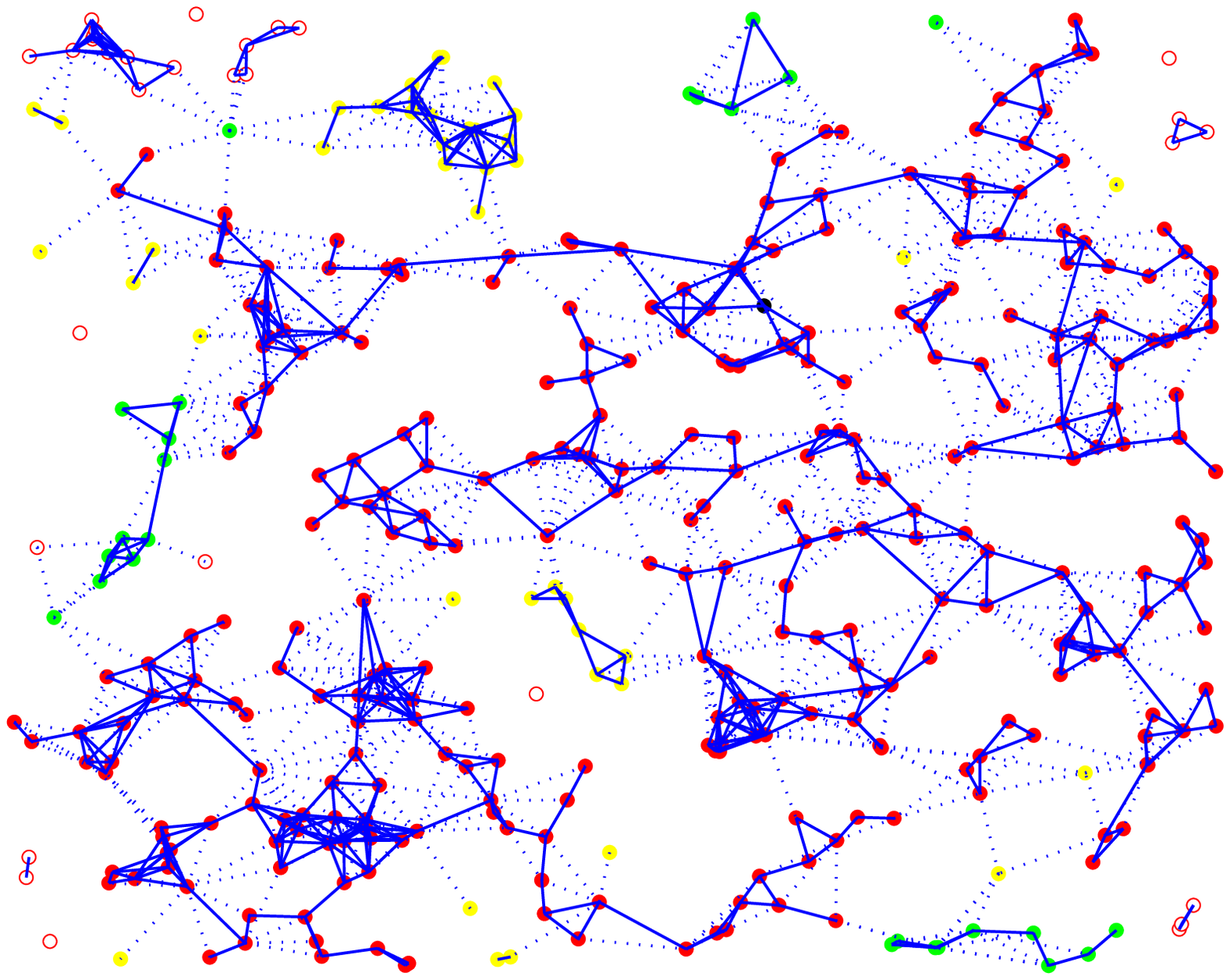}}}
\caption{Percolation in directed SINR graphs without interference where transmission
powers have a binary distribution such that $\mbox{Pr}\{R_i=1\}=0.8$ and
$\mbox{Pr}\{R_i=2\}=0.2$: (a) $\lambda=0.75$, (b) $\lambda=1$.}\label{fig:BDRGG2}
\end{figure}

In this section, we present some simulation results on percolation in directed SINR
graphs. Figure~\ref{fig:BDRGG1} and Figure~\ref{fig:BDRGG2} illustrate percolation
processes in $\overrightarrow{G}(\mathcal{H}_{\lambda},(a,b))$ where the transmission
powers have a binary distribution (see Section III-B).

\begin{figure}[t!]
\centering
\includegraphics[width=3.2in]{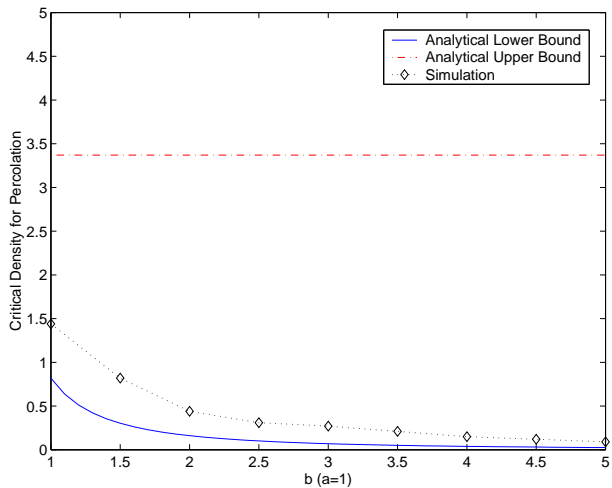}
\caption{Critical density for directed SINR graphs without interference where
transmission powers have a binary distribution such that $\mbox{Pr}\{R_i=1\}=0.5$ and
$\mbox{Pr}\{R_i=b'\}=0.5$ where $b'$ ranges over
$[1,5]$.}\label{fig:BDRGGCriticalDensity}
\end{figure}

Figure~\ref{fig:BDRGGCriticalDensity} shows numerical and simulation results for the
critical density $\overrightarrow{\lambda}_c(a,b)$. Note that the lower bound is quite
tight.

\begin{figure}[t!]
\centerline{ \subfigure[subcritical]{
\includegraphics[width=3.0in]{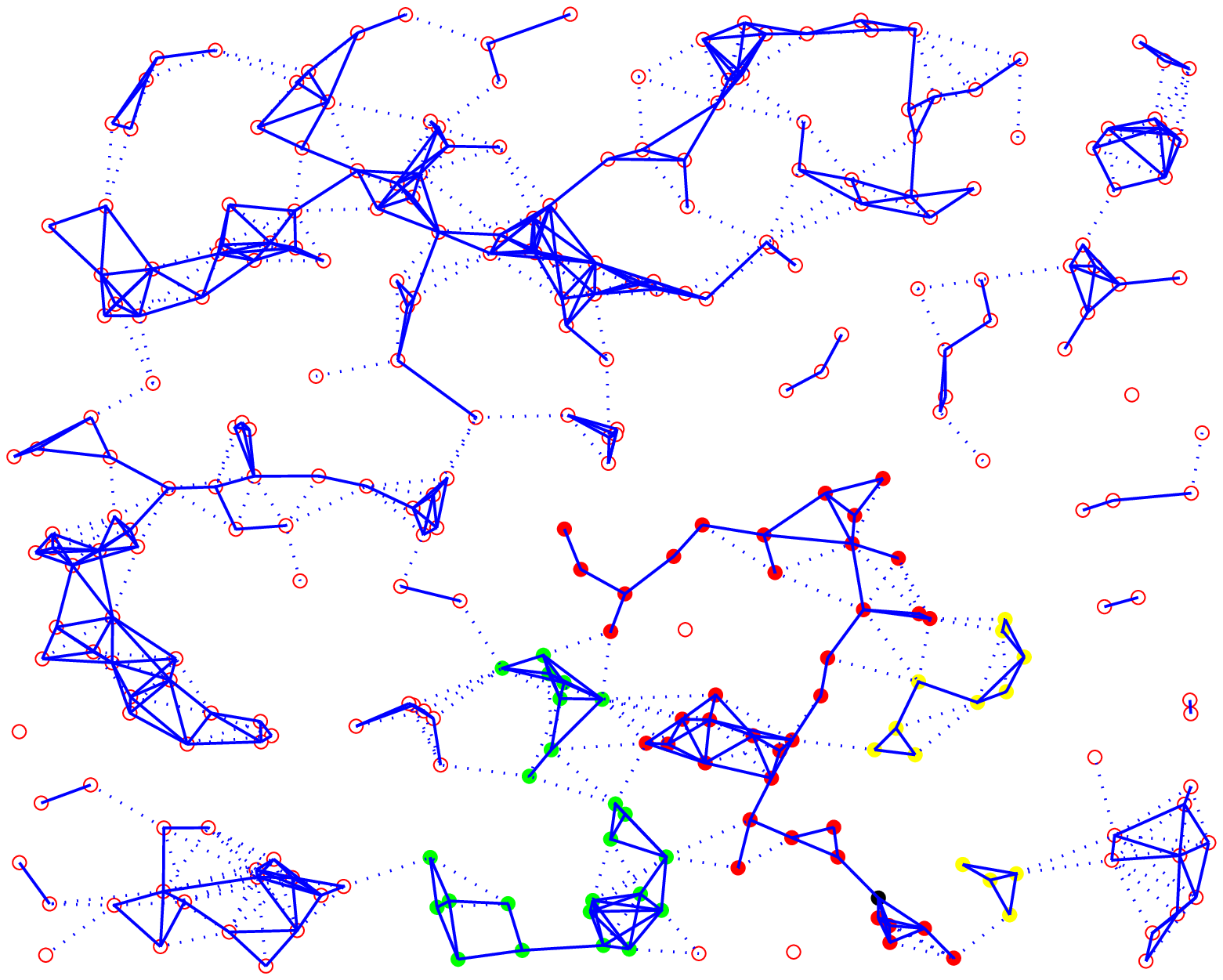}}\hfil
\subfigure[supercritical]{
\includegraphics[width=3.0in]{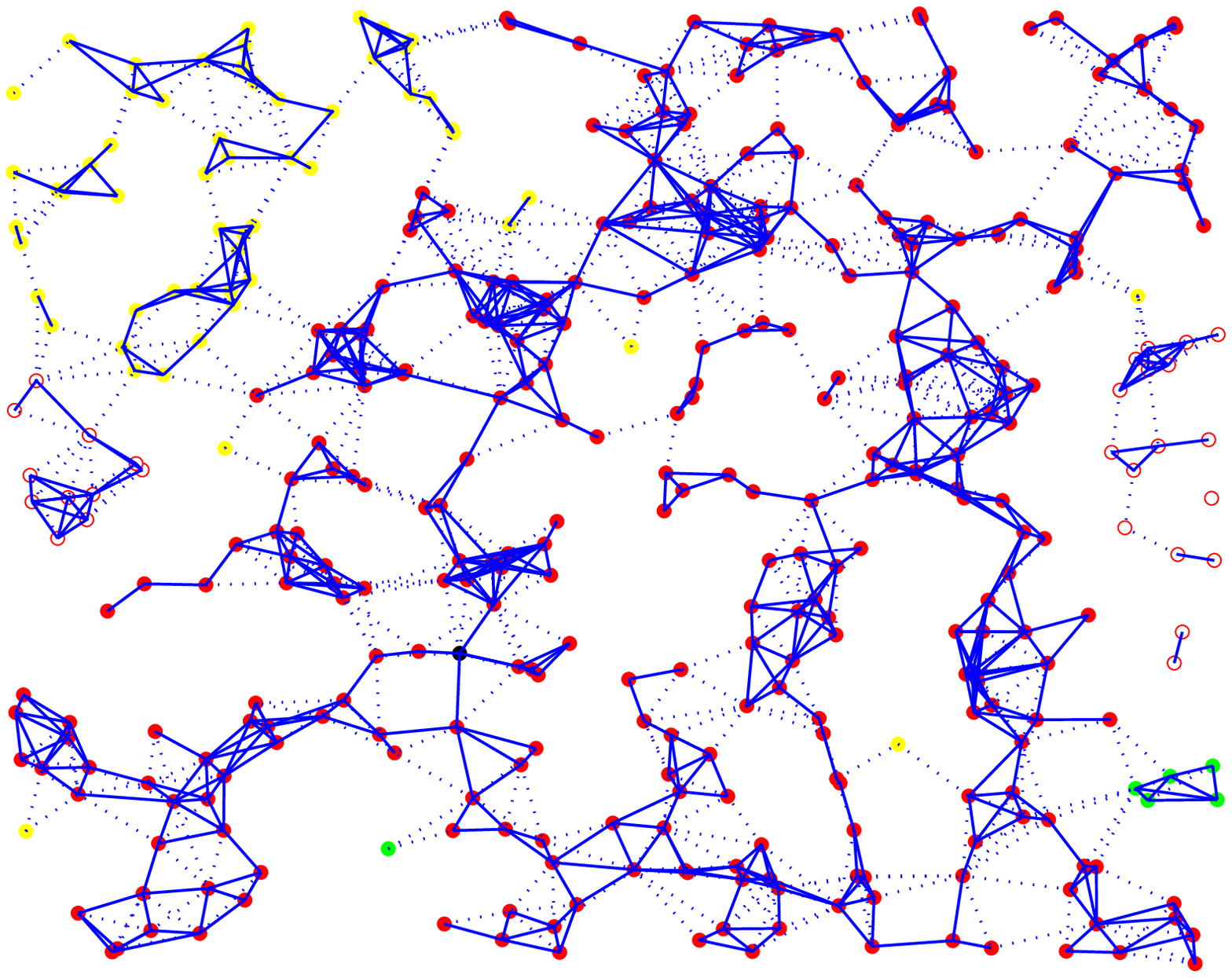}}}
\caption{Percolation in directed SINR graphs without interference where transmission
radii have a power law distribution: $f_R(r)=cr^{-\alpha}$, $r\in(1,2)$ with $\alpha=3$:
(a) $\lambda=0.75$, (b) $\lambda=1$.}\label{fig:DRGG}
\end{figure}

In Figure~\ref{fig:DRGG}, the transmission radius of
$\overrightarrow{G}(\mathcal{H}^{(d)}_{\lambda},\mathcal{P})$ has a power law
distribution as $f_R(r)=cr^{-\alpha}, r\in(1,2)$, where $\alpha=3$ and $c$ is a
normalizing factor.

\begin{figure}[t!]
\centerline{ \subfigure[subcritical]{
\includegraphics[width=3.0in]{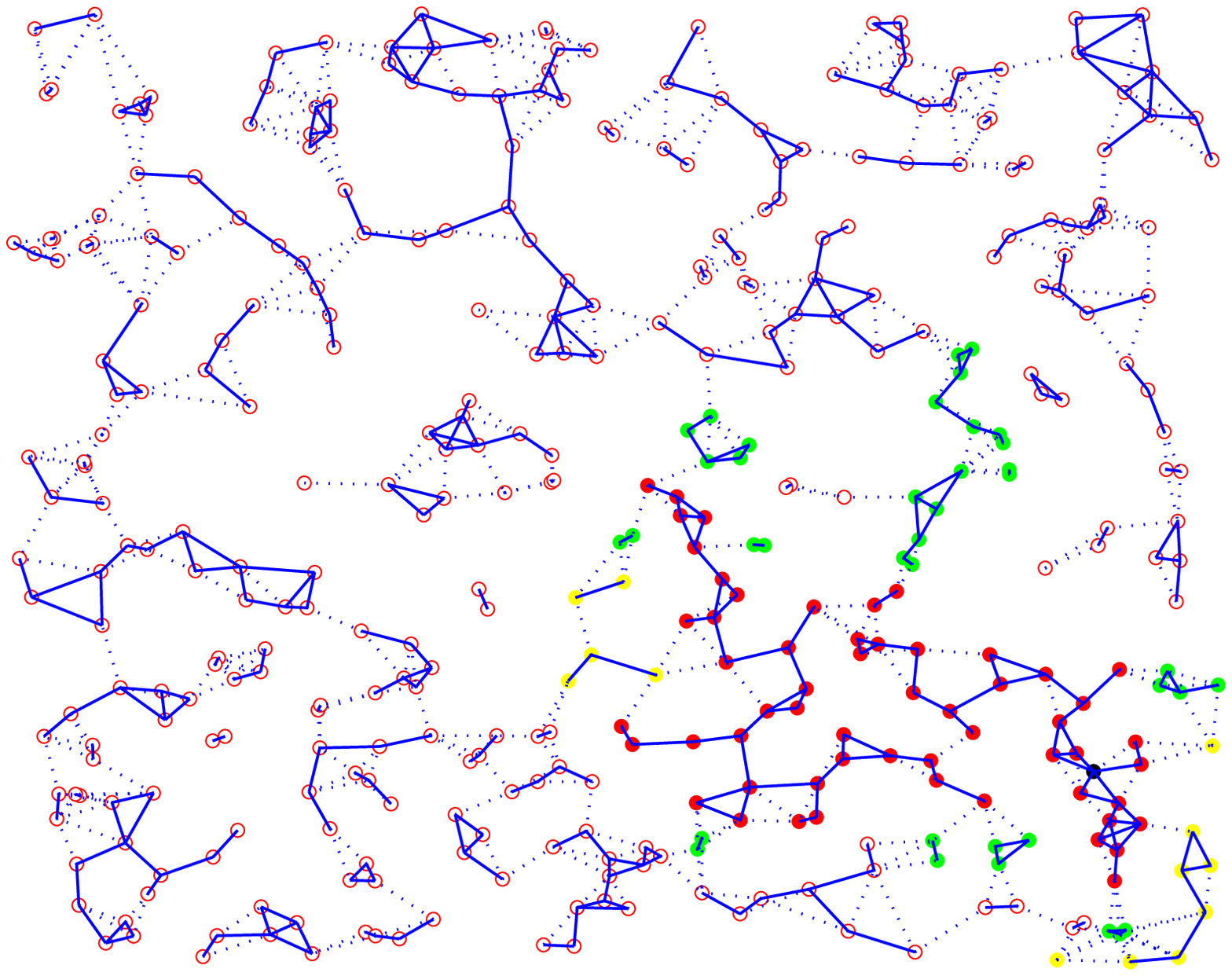}}\hfil
\subfigure[supercritical]{
\includegraphics[width=3.0in]{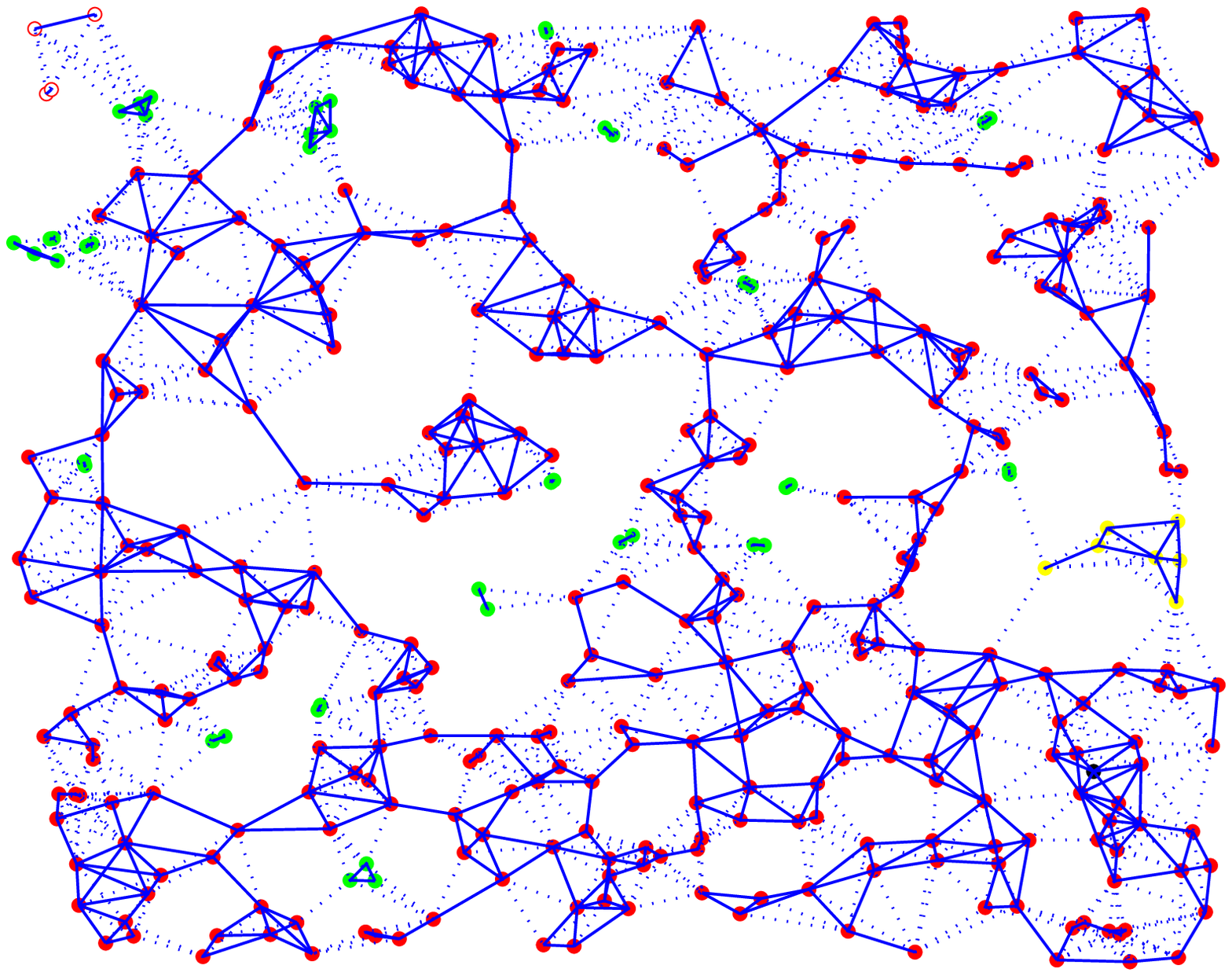}}}
\caption{Percolation in directed SINR graphs where the power of each node has a uniform
distribution over $[p_{min}=1,p_{max}=2]$. The background noise power is $N_0=0.1$. The
successful decoding threshold is $\beta=0.25$. The path-loss function is
$L(d_{ij})=(d_{ij}+(2N_0\beta)^{-1/3})^{-3}$, and the node density is $\lambda=4$: (a)
$\gamma=0.25$, (b) $\gamma=0.1$.}\label{fig:DSINR}
\end{figure}

In Figure~\ref{fig:DSINR}, simulation results of percolation in directed SINR graphs
$\overrightarrow{G}(\mathcal{H}_{\lambda},\mathcal{P},\gamma)$ with interference are
shown. In this case, the transmission power at each node has a uniform distribution over
$[p_{min}=1,p_{max}=2]$, the background noise power is $N_0=0.1$ and the successful
decoding threshold is $\beta=0.25$. The path-loss function is
\[
L(d_{ij})=\left(d_{ij}+\frac{1}{\sqrt[3]{2N_0\beta}}\right)^{-3}
\]
which satisfies our assumptions on
$L(\cdot)$ (i)--(iii).

In all these figures, a randomly picked source node $u$ is represented by a black solid
circle. The nodes in $W_{strong}(u)$ are represented by red solid circles. The nodes in
$W_{in}(u)\backslash W_{strong}(u)$ are represented by yellow solid circles. The nodes in
$W_{out}(u)\backslash W_{strong}(u)$ are represented by green solid circles, and those
nodes not connected to $u$ in anyway are represented by red empty circles. In all of
these cases, we see that phase transitions with respect to the in-component,
out-component, weakly connected component and strongly connected component do take place
at the same time.

\begin{figure}[t]
\centering
\includegraphics[width=3.2in]{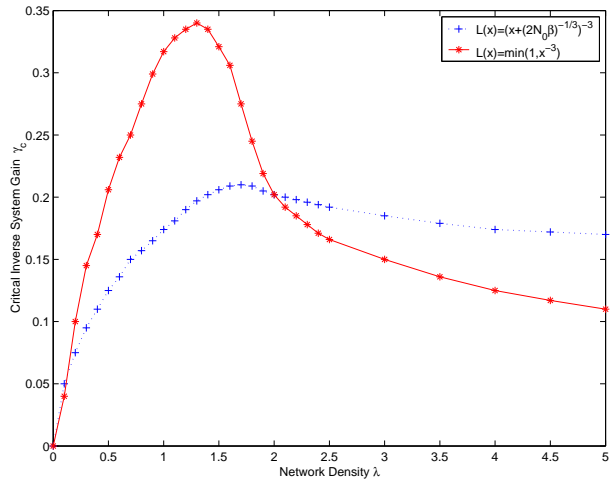}
\caption{Critical inverse system gain for percolation in directed SINR graphs as a
function of node density, for two path loss functions.}\label{fig:DrawGamma}
\end{figure}

The values of the critical inverse system gain $\overrightarrow{\gamma}_c(\lambda)$
depends on the node density and path-loss function. Figure~\ref{fig:DrawGamma} shows
simulation results for $\overrightarrow{\gamma}_c(\lambda)$ with node densities ranging
from 0 to 5, and with two different path-loss functions. In the simulation for
$\overrightarrow{G}(\mathcal{H}_{\lambda},P,\gamma)$, $P$ has a uniform distribution over
$[p_{min},p_{max}]$ with $p_{min}=1$ and $p_{max}=2$.

\section{Conclusions}

In order to understand how the directional nature of wireless communication links affects
global connectivity, we investigated percolation processes in wireless networks modelled
by directed SINR graphs. We first studied interference-free networks ($\gamma=0$), where
we defined four types of phase transitions and showed that they take place at the same
time. By coupling the directed SINR graph with two other undirected SINR graphs and by
using cluster coefficient and re-normalization methods, we further obtained analytical
upper and lower bounds on the critical density $\overrightarrow{\lambda_c}(\mathcal{P})$.
Finally, we showed that with interference ($\gamma>0$), percolation in directed SINR
graphs depends not only on the density but also on the inverse system processing gain
$\gamma$, and there exists a positive and finite critical value
$\overrightarrow{\gamma_c}(\lambda)$, such that the network is percolated only if
$\lambda>\overrightarrow{\lambda_c}(\mathcal{P})$ and
$\gamma>\overrightarrow{\gamma_c}(\lambda)$. We obtained new upper and lower bounds on
$\overrightarrow{\gamma_c}(\lambda)$.

\section*{Appendix}

\subsection{Calculation of $C_3'$}

Let the transmission radii of nodes $i$, $j$ and $k$ be $r_i$, $r_j$ and $r_k$
respectively, i.e.,
\[
r_i = L^{-1}\left(\frac{\beta N_0}{P_i}\right),\qquad r_j = L^{-1}\left(\frac{\beta
N_0}{P_j}\right), \qquad\mbox{and}\qquad r_k = L^{-1}\left(\frac{\beta N_0}{P_k}\right).
\]

Without loss of generality, we assume that $r_j\geq r_i$. Denote the coverage area of
node $i$ located at $\mathbf{x_i}$ with radius $r_i$ by $\mathcal{A}({\mb x_i},r_i)$. The
cluster coefficient is equal to the conditional probability that nodes $i$ and $j$ are
adjacent given they are both adjacent to node $k$. For different ordering on the values
of $r_i$, $r_j$ and $r_k$, the cluster coefficient of
$G'(\mathcal{H}_{\lambda},\mathcal{P})$ is different. In the following, we categorize all
the possibilities into three scenarios: $r_k\geq r_j\geq r_i$, $r_j\geq r_k\geq r_i$, and
$r_j\geq r_i\geq r_k$, and calculate the conditional cluster coefficient separately.

\textbf{\emph{Case I:}} $r_k\geq r_j\geq r_i$

In this case, to determine the cluster coefficient, assuming both nodes $i$ and $j$ are
adjacent to node $k$, the probability that nodes $i$ and $j$ are also adjacent is equal
to the probability that two randomly chosen points in a circle with radius $r_k$ is less
than or equal to a distance $r_j$ apart. That is, the probability that there is a link
between nodes $i$ and $j$ is equal to the fraction of ${\cal A}({\mb x}_j,r_j)$ that
intersects ${\cal A}({\mb x}_k,r_k)$ over ${\cal A}({\mb x}_k,r_k)$.

This fraction is
\[
b(x,y) = \frac{|{\cal A}({\mb x}_k,r_k)\cap{\cal A}({\mb x}_j,r_j)|}{|{\cal A}({\mb
x}_k,r_k)|}.
\]
By averaging over all points in ${\cal A}({\mb x}_k,r_k)$, we obtain the cluster
coefficient for this case:
\begin{eqnarray*}
C_{(r_k\geq r_j\geq r_i)}' &=& \frac{1}{|\mathcal{A}({\mb
x}_k,r_k)|}\int_{{\cal A}({\mb x}_k,r_k)}b(x,y)dx dy\\
 &=&\frac{1}{\pi r_k^2}\int_{{\cal A}({\mb x}_k,r_k)}b(x,y)dx dy.
\end{eqnarray*}
Changing to polar coordinates, we obtain
\[
C_{(r_k\geq r_j\geq r_i)}' = \frac{1}{\pi r_k^2}\int_0^{r_k}\int_0^{2\pi}b(h)hd\psi dh.
\]

\begin{figure}[t]
\centering
\includegraphics[width=2.4in]{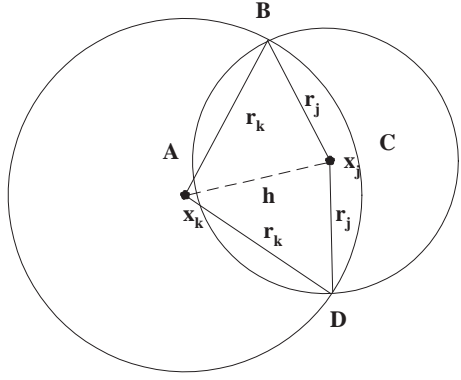}
\caption{Calculation of $C_{(r_k\geq r_j\geq r_i)}'$.}
\end{figure}

As shown in Figure 10, $b(h)$ can be calculated as
\begin{equation}\label{b-h}
b(h)=\frac{\phi_1 r_j^2 +\theta_1 r_k^2-hr_j\sin\phi_1}{\pi r_k^2},
\end{equation}
where
\[
\phi_1\triangleq\angle x_kx_iB=\cos^{-1}\left(\frac{h^2+r_j^2-r_k^2}{2hr_j}\right),
\]
and
\[
\theta_1\triangleq\angle x_ix_kB=\cos^{-1}\left(\frac{h^2+r_k^2-r_j^2}{2hr_k}\right).
\]
That is because the area of intersection is given by
\begin{eqnarray*}
&& S_{x_k\stackrel{\frown}{BCD}}+S_{x_j\stackrel{\frown}{BAD}}
-S_{Bx_kDx_j}\\
& = & \pi r_k^2\cdot\frac{2\theta_1}{2\pi}+\pi
r_j^2\cdot\frac{2\phi_1}{2\pi}-2\cdot\frac{1}{2}hr_j\sin\phi_1\\
& = & \phi_1 r_j^2 +\theta_1 r_k^2-hr_j\sin\phi_1.
\end{eqnarray*}
Note that $b(h)$ is independent of $\psi$. Hence,
\begin{eqnarray*}\label{C-1-prime}
C_{(r_k\geq r_j\geq r_i)}' & = & \frac{1}{\pi r_k^2}\int_0^{r_k}\int_0^{2\pi}b(h)hd\psi dh\nonumber\\
& = & \frac{2}{r_k^2}\int_0^{r_k}b(h)hdh\nonumber\\
& = &\frac{2}{\pi r_k^4}\int_{0}^{r_k}(\phi_1 r_j^2 +\theta_1 r_k^2-hr_j\sin\phi_1)hdh.
\end{eqnarray*}

\textbf{\emph{Case II:}} $r_j\geq r_k\geq r_i$

This scenario is similar to the first case, with  the roles of $r_j$ and $r_k$ exchanged.
Thus, we have
\[
C_{(r_j\geq r_k\geq r_i)}'=\frac{2}{\pi r_j^4}\int_{0}^{r_j}(\theta_1 r_k^2 +\phi_1
r_j^2-hr_k\sin\theta_1)hdh.
\]

\textbf{\emph{Case III:}} $r_j\geq r_i\geq r_k$

In this case, the probability that there is a link between nodes $i$ and $j$ is equal to
the fraction of ${\cal A}({\mb x}_j,r_j)$ that intersects ${\cal A}({\mb x}_k,r_i)$ over
${\cal A}({\mb x}_j,r_j)$.

This fraction is
\[
b(x,y) = \frac{|{\cal A}({\mb x}_j,r_j)\cap{\cal A}({\mb x}_k,r_i)|}{|{\cal A}({\mb
x}_j,r_j)|}.
\]
By averaging over all points in ${\cal A}({\mb x}_j,r_j)$ and changing to polar
coordinates, we obtain the cluster coefficient for this case as
\[
C_{(r_j\geq r_i\geq r_k)}' = \frac{1}{\pi r_j^2}\int_0^{r_j}\int_0^{2\pi}b(h)hd\psi dh.
\]

As in Case I, we can calculate
\[
b(h)=\frac{\phi_2 r_i^2 +\theta_2 r_j^2-hr_j\sin\theta_2}{\pi r_j^2},
\]
where
\[
\phi_2\triangleq\angle x_jx_kB=\cos^{-1}\left(\frac{h^2+r_i^2-r_j^2}{2hr_i}\right),
\]
and
\[
\theta_2\triangleq\angle x_kx_jB=\cos^{-1}\left(\frac{h^2+r_j^2-r_i^2}{2hr_j}\right).
\]

Hence,
\begin{eqnarray*}
C_{(r_j\geq r_i\geq r_k)}' & = & \frac{1}{\pi r_j^2}\int_0^{r_j}\int_0^{2\pi}b(h)hd\psi dh\nonumber\\
& = & \frac{2}{r_j^2}\int_0^{r_j}b(h)hdh\nonumber\\
& = &\frac{2}{\pi r_j^4}\int_{0}^{r_j}(\phi_2 r_i^2 +\theta_2 r_j^2-hr_j\sin\theta_2)hdh.
\end{eqnarray*}

All of the above results are based on the assumption that $r_j\geq r_i$. Since nodes $i$
and $j$ are equivalent, the results also hold when $r_i\geq r_j$. Therefore, the cluster
coefficient for $G'(\mathcal{H}_{\lambda}^{(2)},R)$ can be calculated as
\[
\begin{array}{lll}
C_3'(\mathcal{P}) & = & 2\cdot \displaystyle\left(
\int_{\underline{r}}^{\bar{r}}\int_{r_i}^{\bar{r}}\int_{r_j}^{\bar{r}}
C_{(r_k\geq r_j\geq r_i)}'f_R(r_k)f_R(r_j)f_R(r_i)dr_kdr_jdr_i\right.+\nonumber\\
&&\displaystyle\left.\int_{\underline{r}}^{\bar{r}}\int_{r_i}^{\bar{r}}\int_{r_k}^{\bar{r}}
C_{(r_j\geq r_k\geq r_i)}'f_R(r_j)f_R(r_k)f_R(r_i)dr_jdr_kdr_i\right.+\nonumber\\
&&\displaystyle\left.\int_{\underline{r}}^{\bar{r}}\int_{r_k}^{\bar{r}}\int_{r_i}^{\bar{r}}
C_{(r_j\geq r_i\geq r_k)}'f_R(r_j)f_R(r_i)f_R(r_k)dr_jdr_idr_k\right).
\end{array}
\]

\subsection{Calculation of $\bar{C}$}

To calculate $\bar{C}$, assume both nodes $i$ and $j$ lie within the transmission range
of node $k$. We calculate the probability that nodes $i$ and $j$ are also adjacent. Since
the roles of nodes $i$ and $j$ are interchangeable, we further assume that $r_j\geq r_i$.

It is clear that $C(r_k=a,r_j=a,r_i=a)=C(r_k=b,r_j=b,r_i=b)=C$, where
$C=1-\frac{3\sqrt{3}}{4\pi}$ is the cluster coefficient for random geometric graphs with
identical radii \cite{DaCh02, KoYe07-1, KoYe07-2}. It is also easy to check that
$C(r_k=b,r_j=b,r_i=a)=C(r_k=a,r_j=b,r_i=b)=C$. Therefore we need to consider only two
cases: $r_k=b,r_j=a,r_i=a$ and $r_k=a,r_j=b,r_i=a$.

The first case corresponds to Case I in Appendix A. Thus we have
\[
C(r_k=b,r_j=a,r_i=a)=\frac{2}{\pi b^2}\int_{0}^{b}(\phi_1 a^2+\theta_1 b^2-hb\sin
\theta_1)hdh,
\]
where
\[
\phi_1\triangleq\angle x_kx_iB = \cos^{-1}\left(\frac{h^2+a^2-b^2}{2ah}\right),
\]
and
\[
\theta_1\triangleq\angle x_ix_kB = \cos^{-1}\left(\frac{h^2+b^2-a^2}{2bh}\right).
\]

The latter case corresponds to Case II in Appendix A, thus we have
\[
C(r_k=a,r_j=b,r_i=a)=\frac{2}{\pi b^4}\int_{0}^b(\phi_1 b^2+\theta_1 a^2-ha\sin
\theta_1)hdh.
\]

All the above results are based on the assumption that $r_j\geq r_i$. Since nodes $i$ and
$j$ are equivalent, the results also hold when $r_i\geq r_j$. Therefore, the cluster
coefficient for this model is
\[
\bar{C} =(p_b^3+p_a^3+3p_b^2p_a)C + p_bp_a^2\left(\frac{2}{\pi b^4}\right)\cdot\int_0^b
\left[(\phi_1+\theta_1)(a^2+b^2)+h\sin \theta_1 (a+b)\right]hdh.
\]

\bibliography{PercolationTopic2}
\bibliographystyle{ieeetr}

\end{document}